# Toward Quantum-Limited Position Measurements Using Optically Levitated Microspheres

KENNETH G. LIBBRECHT AND ERIC D. BLACK[1]

*Norman Bridge Laboratory of Physics, California Institute of Technology 264-33, Pasadena, CA 91125*

**Abstract.** We describe the use of optically levitated microspheres as test masses in experiments aimed at reaching and potentially exceeding the standard quantum limit for position measurements. Optically levitated microspheres have low mass and are essentially free of suspension thermal noise, making them well suited for reaching the quantum regime. Table-top experiments using microspheres can bridge the gap between quantum-limited position measurements of single atoms and measurements with multi-kg test masses like those being used in interferometric gravitational wave detectors.

## 1. Introduction

There is a rich literature surrounding the quantum limits inherent in the precise measurement of the position of a test mass, as well as techniques for potentially exceeding these limits [1-3]. The range of masses involved in such measurements spans some 27 orders of magnitude, from individual laser-cooled atoms [4,5] to test masses weighing tens of kilograms [6,7]. Position measurements of the latter are of particular interest for the detection of gravitational radiation from astrophysical sources, and it is expected that the next generation of interferometric gravitational-wave detectors will be largely quantum limited over a broad frequency range [8].

If we restrict ourselves to optical detection of macroscopic test masses, then one typically speaks of position measurements using laser interferometry, where the test mass is one of the interferometer mirrors. In this case, a good figure-of-merit for an experiment aimed at reaching the quantum regime is the ratio $F(\omega) = \delta x_{meas}(\omega)/\delta x_{quant}(\omega)$, where $\delta x_{meas}(\omega)$ is the measured displacement noise spectral density at the measurement angular frequency $\omega$, and $\delta x_{quant}(\omega) = (\hbar/m\omega^2)^{1/2}$, where $m$ is the mirror mass [3]. Reaching the quantum regime, i.e. achieving $F \approx 1$ at any measurement frequency, has proven to be a surprisingly elusive goal, with the best recent experiments only attaining $F \sim 10^3$ [3,9]. New proposed experiments, using both low-mass cantilevers [10,11] and advanced gravitational-wave detectors [8], show promise for achieving $F \approx 1$.

We show here that optically levitated microspheres provide another promising route to realizing quantum-limited position measurements. Because suspension thermal noise is virtually absent from a levitated microsphere, it appears possible to reach the quantum regime with a relatively simple shadow sensor measurement scheme using the levitating laser beam.

## 2. Noise Analysis

Consider the case of a dielectric sphere of radius $R$ held at the focus of a vertical laser beam of total power $P$ and waist radius $w$, as shown in Figure 1. Since the sphere scatters light into large angles, we can determine its transverse position by the shadow imposed on an image of the laser beam waist. From a straightforward analysis of this shadow detection scheme, we find that shot noise in the measurement of transverse position produces a noise spectral density of

$$\delta x_{shot}(\omega) \approx G\sqrt{\frac{\hbar c R^2}{\lambda P}}$$

$$\approx 5\times 10^{-16} G \left(\frac{R}{1\,\mu\text{m}}\right)\left(\frac{100\,\text{mW}}{P}\right)^{1/2}\left(\frac{1\,\mu\text{m}}{\lambda}\right)^{1/2} \text{m}/\sqrt{\text{Hz}}$$

where $\lambda < R$ is the laser wavelength, and we have assumed the optimal case of $w \approx R$. The geometrical factor $G$ depends on $R/\lambda$, $R/w$, and the precise details of the Mie scattering by the sphere, as well as the photodetection efficiency. A full analysis of $G$ is beyond the scope of this paper, but we estimate that a realistic experiment might achieve $G \approx 10$ or perhaps lower.

---

[1]Address correspondence to *kgl@caltech.edu*; URL: http://www.its.caltech.edu/atomic/

Since quantum noise for this system is

$$\delta x_{quant}(\omega) \approx 2\times 10^{-13}\left(\frac{1\,\mu\text{m}}{R}\right)^{3/2}\left(\frac{2\pi \cdot 100\,\text{Hz}}{\omega}\right) \text{m}/\sqrt{\text{Hz}}$$

we see that it appears feasible to achieve $F < 1$ in a realistic experimental system using few-micron-diameter spheres. To achieve this goal, however, we must also reduce numerous other noise contributions to below $\delta x_{quant}(\omega)$. These include the following.

**Residual gas damping**. This can be expressed as a damping time for a moving sphere [6], which in our case is

$$\tau_{gasdamp} \approx 7\left(\frac{R}{1\,\mu\text{m}}\right)\left(\frac{10^{-10}\,\text{Torr}}{P_{vac}}\right) \text{years}$$

where $P_{vac}$ is the vacuum pressure. The noise from gas damping is below the radiation pressure noise if $\tau_{gasdamp}$ is longer than the damping time that derives from radiation pressure acting on the sphere [12], which in our case is

$$\tau_{raddamp} \approx 10^{-3}\left(\frac{R}{1\,\mu\text{m}}\right)\left(\frac{10^{10}\,\text{W/m}^2}{I}\right) \text{years}$$

where $I \approx P/4w^2$ is the central intensity of the incident laser beam. Clearly the condition $\tau_{raddamp} > \tau_{gasdamp}$ can be realized using a standard ultra-high-vacuum system.

We note that light forces on the sphere produce a dipole trap with transverse frequency (for $w \approx R$) of

$$\omega_{trap} \approx \sqrt{\frac{I}{\rho R^2 c}}$$

$$\omega_{trap}/2\pi \approx 5\left(\frac{1\,\mu\text{m}}{R}\right)\left(\frac{I}{10^9\,\text{W/m}^2}\right)^{1/2} \text{kHz}$$

so the mechanical $Q$ of the trapped sphere arising from radiation damping is approximately

$$Q \approx 10^{10}\left(\frac{10^9\,\text{W/m}^2}{I}\right)^{1/2}$$

independent of $R$. The timescale for measuring this $Q$ is of order $\tau_{raddamp}$, and such a measurement would be the first of its kind to observe this form of radiation damping.

We also note that launching and trapping dielectric spheres in ultra-high-vacuum has been demonstrated using ultrasonic launching coupled with optical position detection and feedback damping of charged spheres [13]. Similar techniques with somewhat smaller spheres would be the first step in the experiments we are proposing. The damping times above are sufficiently long that all motional degrees of freedom must initially be damped via active feedback control.

**Laser jitter noise**. This could be reduced to manageable levels by mounting critical optical components on seismically and acoustically isolated structures. The laser beam motion could be stabilized using a high-finesse, suspended, fixed-length mode cleaning cavity. Any residual laser motion could be measured down to the shot noise level by the photodetection system (see Figure 1), which itself would be on a suspended platform. In practice, we believe these techniques could reduce the laser jitter noise to acceptable levels at frequencies above ~ 50 Hz. Below 50 Hz, we expect the noise in any practical experiment would rise precipitously because of residual seismic noise, as is seen in suspended interferometers used in gravitational-wave detection [6,7].

## 3. Discussion

The system described above is inherently lossy, in that many photons striking the test mass are not efficiently used to detect its position. These photons increase radiation pressure noise without improving the position measurement. Nevertheless, the experiment outlined above should allow an observation of the main spectral signature of a quantum-limited optical position measurement, namely the appearance of both shot noise and radiation-pressure noise in a single measured noise spectrum $\delta x_{meas}(\omega)$ [8].

It may be possible to substantially reduce these losses using optically levitated test masses in the shape of small platelets, which might be levitated with a fixed orientation and coated *in situ* with a high-reflectivity coating. One could then envision a small optically levitated mass acting as one end of a high-finesse optical cavity, perhaps with a finesse of $10^4$ or higher. Such a system would allow the realization of a wealth of experiments demonstrating optical bistability [14] and the quantum entanglement of optical and positional degrees of freedom [15,16].

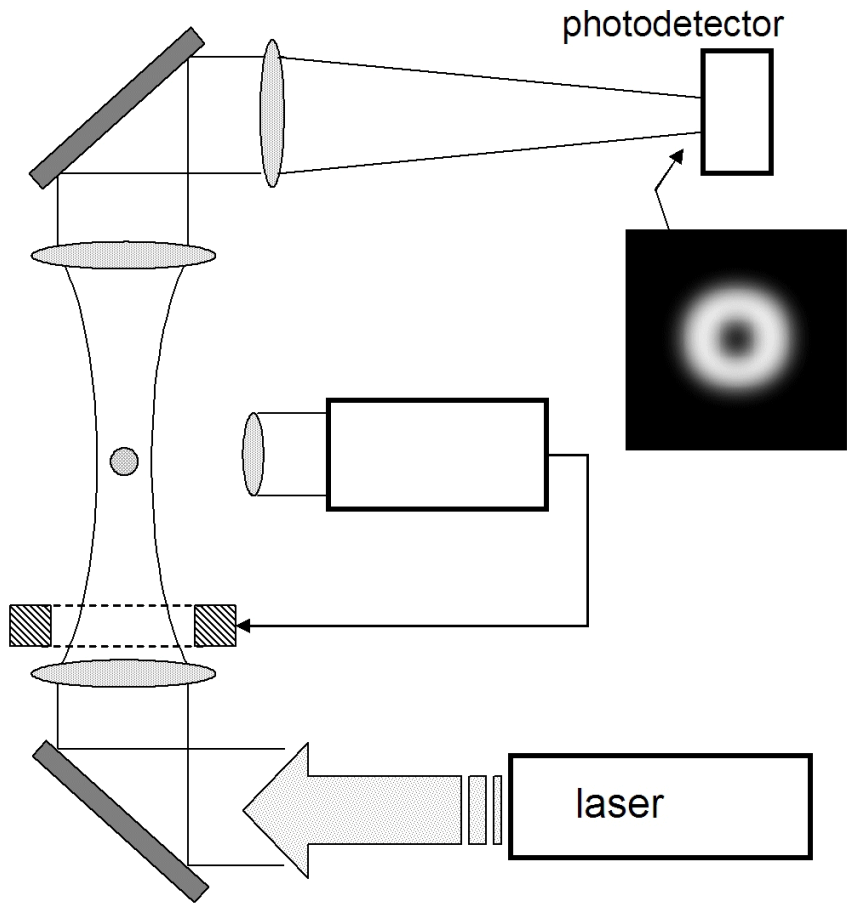

Figure 1. This shows one possible experimental arrangement for achieving a quantum-limited position measurement. A charged dielectric microsphere is held at the focus of a vertically propagating laser beam, where light forces confine the sphere in the horizontal plane. The sphere's vertical position is measured optically (using scattered laser light) and controlled by applying a voltage to a ring electrode beneath the sphere. The image at the photodetector shows the laser waist with a partial shadow from the microsphere.